\documentclass[aps,prl,twocolumn,groupedaddress]{revtex4} 

\usepackage{amssymb,amsfonts,amsmath}
\usepackage{graphicx}
\usepackage{verbatim}

\newcommand{\beqn}{\begin{eqnarray}}
\newcommand{\eeqn}{\end{eqnarray}}
\newcommand{\beq}{\begin{equation}}
\newcommand{\eeq}{\end{equation}}
\newcommand{\bra}[1]{\langle{#1}|}
\newcommand{\ket}[1]{|{#1}\rangle{}}
\newcommand{\bracket}[2]{\langle{#1}|{#2}\rangle{}}
\newcommand{\av}[1]{\langle{#1}\rangle{}}
\newcommand{\mk}[2]{\newcommand{#1}{#2}}
\newcommand{\rmk}[2]{\renewcommand{#1}{#2}}

\mk{\bbn}{b^+_nb^-_n}
\mk{\bbm}{b^+_mb^-_m}
\mk{\bbnn}{b^+_nb^-_n(n)}
\mk{\bbmn}{b_m^+b_m^-(n)}
\mk{\bb}{\bd_n\bl_n}
\mk{\bbbn}{\bd_n{\bar{b}^-_n}}
\mk{\bbbm}{\bd_m{\bar{b}^-_m}}
\mk{\bmb}{{\bar{\bl_m}}}
\mk{\bG}{{\bf{{G}}}}
\mk{\bmn}{\hat{b}_m^-(n)}
\mk{\bnn}{\hat{b}_n^-(n)}
\mk{\bbarn}{\hat{\bar{b}}_n^-}
\mk{\bbarm}{\hat{\bar{b}}_m^-}
\mk{\betan}{\hat{\beta}^-_n}
\mk{\betam}{\hat{\beta}^-_m}
\mk{\bbar}{\bar{b}^-}
\mk{\bd}{\hat{b}^+}
\mk{\ad}{\hat{a}^+}
\mk{\bl}{\hat{b}^-}
\mk{\al}{\hat{a}^-}
\mk{\qn}{q(n)}
\mk{\gn}{g(n)}
\mk{\qh}{\hat{q}_n}
\mk{\gh}{\hat{g}_n}
\mk{\qb}{\bar{q}}
\mk{\gb}{\bar{g}}
\mk{\Dn}{\hat{\Delta}_n}
\mk{\Gah}{\hat{\Gamma}_n}
\mk{\Lah}{\hat{\Lambda}_n}
\mk{\Deh}{\hat{\Delta}_n}
\mk{\Gjk}{G^{jk}}
\mk{\bj}{\<j|}
\mk{\bk}{\<k|}
\mk{\Gl}{{G_\ell}}
\mk{\Del}{\Delta}

\mk{\ehat}{\bf{\hat{ e}}}
\rmk{\L}{{\cal L}}
\rmk{\H}{\hat{H}}
\mk{\n}{\langle n\rangle}

\begin{document}

\title{A stochastic spectral analysis of transcriptional regulatory cascades}

\author{Aleksandra M. Walczak}
\email[]{awalczak@princeton.edu}
\affiliation{Princeton Center for Theoretical Science, Princeton University, Princeton, NJ 08544}

\affiliation{}
\author{Andrew Mugler}
\email[]{ajm2121@columbia.edu}

\affiliation{Department of Physics, Columbia University, New York, NY 10027}
\author{Chris H. Wiggins}
\email[]{chris.wiggins@columbia.edu}
\affiliation{Department of Applied Physics and Applied Mathematics, Center for Computational Biology and Bioinformatics, Columbia University, New York, NY 10027}

\begin{abstract}

The past decade has seen great advances in our understanding of the role of noise in gene
regulation and the
physical limits to signaling in biological networks. Here we introduce the {\it
spectral} method for computation of the joint probability
distribution over all species in a biological network.  The spectral method
exploits the natural eigenfunctions of the master
equation of birth-death processes to solve for the joint
distribution of modules within the network, which then inform each
other and facilitate calculation of the entire joint distribution. We
illustrate the method on a ubiquitous case in
nature: linear regulatory cascades. The efficiency of the method 
makes possible numerical optimization of the
input and regulatory parameters, revealing design properties of, e.g., the
{\it most informative} cascades.
We find, for threshold
regulation, that a cascade of strong regulations converts a unimodal
input to a bimodal output, that multimodal
inputs are no more informative than bimodal inputs, and that a chain of
up-regulations outperforms a chain of down-regulations.  We anticipate
that this numerical
approach may be useful for modeling noise in a variety of small
network topologies in biology.

\end{abstract}

\maketitle

Transcriptional regulatory networks
are composed of genes and proteins, which are often
present in small numbers in the cell \cite{Hooshangi, Thattai}, rendering 
deterministic models poor descriptions of
the counts of protein molecules observed experimentally
\cite{Paulsson2, Elowitz, Ozbudak, Swain, Acar, Pedraza, Thattai2}. Probabilistic approaches have proven necessary to account fully for the variability of molecule numbers within a homogenous population of cells. 
A full stochastic description of even a small regulatory network proves quite challenging. Many efforts have been made to refine simulation approaches \cite{vanZon, vanZon2, Allen, MacNamara, Ushikubo}, which are mainly based on the varying step Monte Carlo or `Gillespie' method \cite{Bortz, Gillespie}. Yet expanding full molecular simulations to larger systems and scanning parameter space is computationally expensive. 
On the other hand the interaction of many protein and gene types makes analytical methods hard to implement. A wide class of approximations to the master equation, which describes the evolution of the probability distribution, focuses on limits of large concentrations or small switches \cite{Paulsson, TanaseNicola, Walczak}. Approximations based on timescale separation of the steps of small signaling cascades have been successfully used to calculate escape properties
\cite{Lan,Lan2,Lan3}.   

In this paper we introduce a new method for calculating the steady-state distributions of chemical 
reactants. 
The procedure, which we call the {\it spectral method}, relies on exploiting the natural basis of a simpler problem from the same class. The full problem is then solved numerically as a an expansion in this basis,
reducing the master equation
to a set of linear algebraic equations. We break up the problem into two parts: a preprocessing step, which can be solved algorithmically; and the parameter-specific step of obtaining the actual probability distributions. The spectral method allows for huge computational gains with respect to simulations.

\begin{figure}
\begin{center}
\includegraphics[scale=0.4]{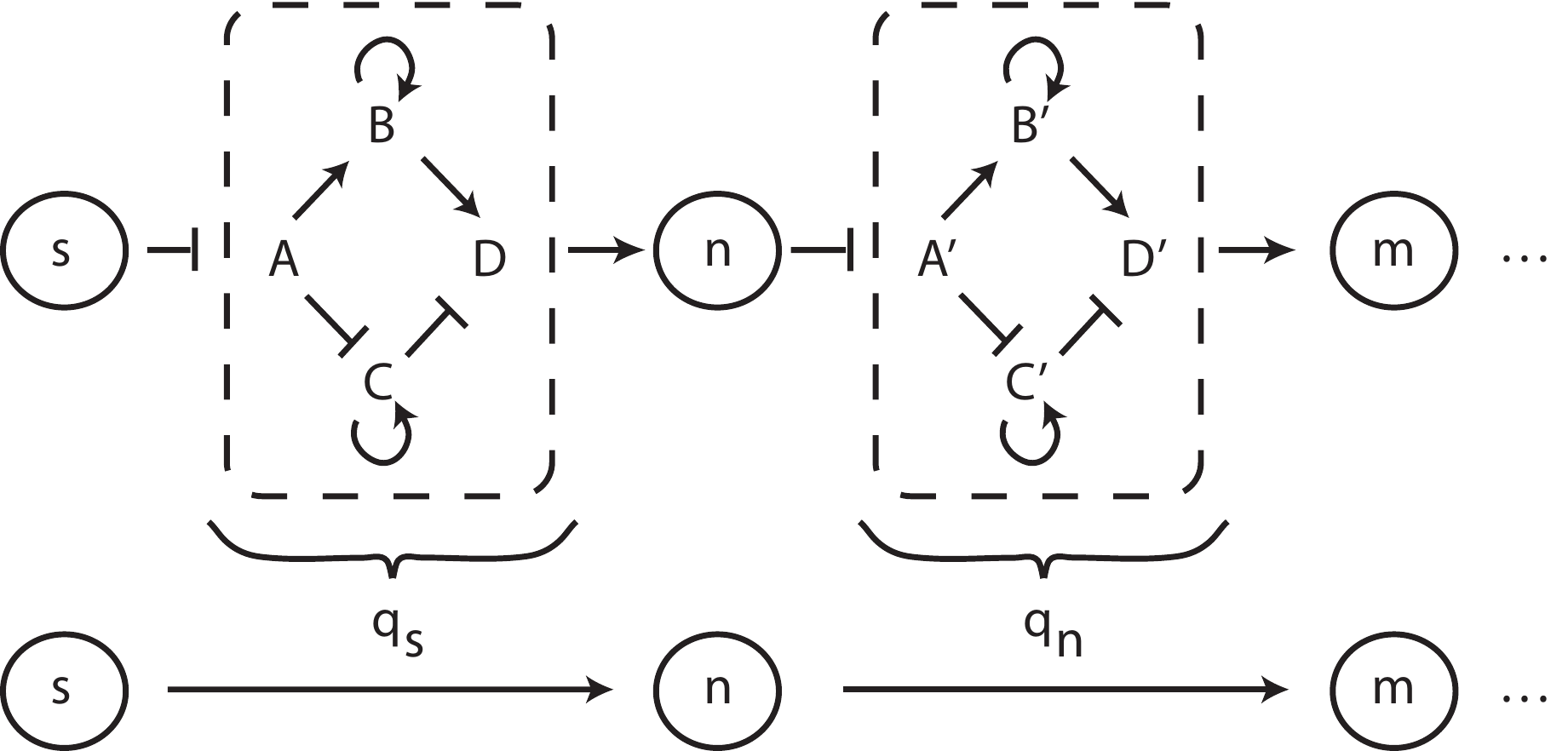}
\caption{A schematic representation of a general signaling cascade.  Interactions between species of interest may include intermediate processes; we take a coarse-grained approach, condensing these intermediate processes into a single effective regulatory function.  For example, the regulatory function $q_n$ describes the creation rate of a species with copy count $m$ as a function of the copy count $n$ of the previous species.}
\label{blackbox}
\end{center}
\end{figure}

We illustrate the spectral method for the case of regulatory cascades: downstream genes responding to concentrations of transcription factors produced by upstream genes which are linked to external cues. 
Cascades play an important role in a diversity of cellular processes \cite{Gomperts, Ting, Detwiler}, from decision making in development \cite{Bolouri} to quorum sensing among cells \cite{Bassler}.
We take a coarse-grained approach, modeling each step of a cascade with a general regulatory function that depends on the copy number of the reactant at the previous step (cf.\ Fig.\ \ref{blackbox}).
While the method as implemented describes arbitrary regulation functions, we optimize the information transmission in the case of the most biologically simple
regulation function: a discontinuous threshold, in which a species is created at a high or low rate depending on the copy count of the species directly upstream.  
In the next sections, we outline the spectral method and present in detail our findings regarding signaling cascades.

\section{Method}

We calculate the steady-state joint distribution
for $L$ chemical species in a cascade (cf. Fig.\ \ref{blackbox}).
The approach we
take involves two key observations: the master equation, being linear,\footnote{A master equation in which the coordinate appears explicitly (e.g.\ through $q_s$ or $q_n$ in Eqn.\ \ref{master3}) is sometimes
mistakenly termed ``nonlinear'' in the literature, perhaps discouraging calculations which exploit
its inherent linear algebraic structure. We remind the reader that this equation is perfectly linear.}
benefits from solution in terms of its eigenfunctions; and the
behavior of a given species should depend only weakly on distant
nodes given the proximal nodes.

The second of these observations can be illustrated succinctly by considering a
three-gene cascade in which the first may be eliminated by marginalization. 
For three species obeying $s\xrightarrow{q_s}n\xrightarrow{q_n}m$ as in Fig.\ \ref{blackbox}, we have the
linear master equation
\beqn
\label{master3}
\dot{p}_{snm}
	=\tilde{\rho}\left[gp_{(s-1)nm}-gp_{snm}+(s+1)p_{(s+1)nm}-sp_{snm}\right]\nonumber\\
	+q_s p_{s(n-1)m}-q_s p_{snm}+(n+1)p_{s(n+1)m}-np_{snm}\nonumber\\
	+\rho\left[q_np_{sn(m-1)}-q_np_{snm}+(m+1)p_{sn(m+1)}-mp_{snm}\right].
\eeqn
Here time is rescaled by the first gene's degradation rate, so that each gene's creation rate ($g$, $q_s$, or $q_n$) is normalized by its respective degradation rate; $\tilde{\rho}$ and $\rho$ are the ratios of the first and third gene's degradation rate to the second's, respectively.

To integrate out the first species, we sum over $s$. We then introduce $g_n$, the effective regulation of $n$, by
\beqn
\label{markov2}
\sum_s q_s p_{nms}
=p_{nm}\sum_sq_sp_{s|nm}
\approx
p_{nm}\sum_s q_s p_{s|n}
\equiv 
g_n p_{nm}.
\eeqn
Here we have made the Markovian approximation that $s$ is conditionally independent of $m$ given $n$.   Generally speaking, the probability distribution
depends on all steps of the cascade.  However since there are no loops in the cascades we consider here, we assume in Eqn.\ \ref{markov2} that at steady-state each species is not affected by species two or more steps downstream in the cascade. The validity of the Markovian approximation is tested using both a non-Markovian tensor implementation of the spectral method and a stochastic simulation using the Gillespie algorithm \cite{Gillespie}, as discussed in {\it Supplementary Material}.  We find that the approximation produces accurate results for all but the most strongly discontinuous regulation functions; even in these cases qualitative features such as modality of the output distribution and locations of the modes are preserved. Armed with the Markovian approximation the equation for the remaining two species simplifies to
\beqn
\label{mastern}
\dot{p}_{nm}=g_{n-1}p_{(n-1)m}-g_np_{nm}+(n+1)p_{(n+1)m}-np_{nm}\nonumber\\
	+\rho\left[q_np_{n(m-1)}-q_np_{nm}+(m+1)p_{n(m+1)}-mp_{nm}\right].
\eeqn
This procedure can be extended indefinitely for a cascade of arbitrary length $L$, in which modules consisting of pairs of adjacent species are each described by two-dimensional master equations.

The distribution for the first two species is obtained by summing
over all other species,
which gives an equation of the same form as Eqn.\ \ref{mastern} but with $g_n=g=$ constant.  If instead the input distribution is an arbitrary $p_n$, the distribution for the first two species is still described by Eqn.\ \ref{mastern}, but with $g_n$ calculated recursively from $p_n$ via $g_n=(-np_n+g_{n-1}p_{n-1}+(n+1)p_{n+1})/p_n$
with $g_0=p_1/p_0$ to initialize.\footnote{The recursion can equivalently be performed in the reverse direction, with $g_N=0$, $g_{N-1}=Np_N/p_{N-1}$, and $g_{n-1}=(g_np_n+np_n-(n+1)p_{n+1})/p_{n-1}$, where $N$ is a cutoff in $n$.  In several test cases we found that reverse recursion is more numerically stable than forward recursion at large $N$.}
Describing the start of a cascade (with arbitrary input distribution) and describing subsequent steps both amount to solving Eqn.\ \ref{mastern} with $g_n$ given by either the recursive equation
or Eqn.\ \ref{markov2} respectively.

We solve Eqn.\ \ref{mastern} by defining the generating function \cite{vanKampen} $G(x,y)=\sum_{n,m}p_{nm}x^ny^m$ over complex variables $x$ and $y$.\footnote{Setting $x=e^{ik_1}$ and $y=e^{ik_2}$ makes clear that the generating function is simply the Fourier transform.}
It will prove more convenient to write the generating function in a state space as $\ket{G}=\sum_{n,m}p_{nm}\ket{n,m}$,\footnote{$G(x,y)$ can be recovered by projecting onto position space $\bra{x,y}$, with $\bracket{x}{n}=x^n$ and $\bracket{y}{m}=y^m$.} with inverse transform $p_{nm}=\bracket{n,m}{G}$, where the states $\ket{i}$ and $\bra{i}$, for $i \in \{n,m\}$, along with the inner product $\bracket{i}{i'}=\delta_{ii'}$, define the protein number basis. With these definitions, Eqn.\ \ref{mastern} at steady-state becomes $0=\H\ket{G},$
where 
\beq
\label{eom}
\H=\bd_n\bnn+\rho\bd_m\bmn.
\eeq
Here
we have introduced raising and lowering operators in protein space  \cite{Mattis, Zeldovich, Doi, Sasai} obeying $\bd_i\ket{i}=\ket{i+1}-\ket{i}$ for $i \in \{n,m\}$, $\bnn\ket{n}=n\ket{n-1}-\gh\ket{n}$ and $\bmn\ket{n,m}=m\ket{n,m-1}-\qh\ket{n,m}$,\footnote{The adjoint operations are $\bra{i} \bd_i=\bra{i-1}-\bra{i}$ for $i \in \{n,m\}$, $\bra{n} \bnn=(n+1)\bra{n+1}-g_n\bra{n}$, and $\bra{n,m} \bmn=(m+1)\bra{n,m+1}-q_n \bra{n,m}$.} and the regulation functions have become operators obeying $\gh\ket{n}=g_n\ket{n}$ and $\qh\ket{n}=q_n\ket{n}$.

Were the operators $\bnn$ and $\bmn$ not $n$-dependent, $\H$ would be easily diagonalizable.  In fact, this corresponds to the uncoupled case, in which there is no regulation, and both upstream and downstream gene undergo independent birth-death processes with Poisson steady-state distributions.  We exploit this fact by working with the respective deviations of $g_n$ and $q_n$ from some constant creation rates $\gb$ and $\qb$.  Then $\H$ can be partitioned as $\H=\H_0+\H_1$, where
\beqn
\H_0&=&\bd_n\bbar_n+\rho\bd_m\bbar_m\\
\H_1&=& \bd_n\Gah + \rho\bd_m\Deh,
\eeqn
and we define new operators $\bbar_n\ket{n}=n\ket{n-1}-\gb\ket{n}$, $\bbar\ket{m}=m\ket{m-1}-\qb\ket{m}$,
$\Gah=\gb-\gh$, and $\Deh=\qb-\qh$.  $\Gah$ and $\Deh$ capture the respective deviations of $g_n$ and $q_n$ from $\gb$ and $\qb$, and $\H_0$ is diagonal in the eigenbases $\ket{j}$ and $\ket{k}$ of the uncoupled birth-death processes at rates $\gb$ and $\qb$ respectively;\footnote{In position space the eigenfunctions are $\bracket{x}{j}=(x-1)^je^{\gb(x-1)}$ and $\bracket{y}{k}=(y-1)^ke^{\qb(y-1)}$.  The operators $\bd$ and $\bbar$ raise and lower in eigenspace: $\bd_n \ket{j} =\ket{j+1}$ and $\bbar_n\ket{j}=j\ket{j-1}$ (or $\bra{j} \bd_n=\bra{j-1}$ and $\bra{j} \bbar_n=(j+1)\bra{j+1}$), and similarly for $n\rightarrow m$ and $j\rightarrow k$.}
specifically $\H_0 \ket{j,k}=(j+\rho k)\ket{j,k}$.  Projecting Eqn.\ \ref{eom} onto the eigenbasis yields the linear equation of motion
\beq
\label{final}
(j+\rho k)\Gjk+\sum_{j'}\Gamma_{j-1,j'}G^{j'k}+\rho\sum_{j'}\Delta_{jj'}G^{j',k-1}=0,
\eeq
where $G^{jk}=\bracket{j,k}{G}$, $\Gamma_{jj'}=\sum_n (\gb-g_n)\bracket{j}{n}\bracket{n}{j'}$, and $\Delta_{jj'}=\sum_n (\qb-q_n)\bracket{j}{n}\bracket{n}{j'}$.   Eqn.\ \ref{final} exploits the subdiagonal nature of the $k$-dependence; it is initialized using $G^{j0}=\sum_np_n\bracket{j}{n}$, then solved exactly by matrix inversion for each subsequent $k$. The joint distribution is retrieved via the inverse transform as
\beq
p_{nm}=\sum_{jk} \bracket{n}{j} G^{jk} \bracket{m}{k}.
\eeq

One computational advantage is that the overlap integrals $\bracket{n}{j}$ and $\bracket{j}{n}$ need only be evaluated explicitly for $\bracket{n}{j=0}=e^{-\gb}\gb^n/n!$ and $\bracket{j}{n=0}=(-\gb)^j/j!$; all other values can be obtained recursively using the selection rules $\bracket{n}{j+1}=\bracket{n-1}{j}-\bracket{n}{j}$ and $\bracket{j}{n+1}=\bracket{j-1}{n}+\bracket{j}{n}$.\footnote{The selection rules are derived by starting with $\bra{n}\bd_n\ket{j}$ or $\bra{j}\bd_n\ket{n}$ and allowing $\bd_n$ to act both to the left and to the right.  Alternatively, one may use $\bbar_n$, obtaining $\bracket{n+1}{j}=(j\bracket{n}{j-1}+\gb\bracket{n}{j})/(n+1)$ and $\bracket{j+1}{n}=(n\bracket{j}{n-1}-\gb\bracket{j}{n})/(j+1)$, initialized with $\bracket{n=0}{j}=(-1)^je^{-\gb}$ and $\bracket{j=0}{n}=1$.  We find the latter relations yield smoother distributions $p_{nm}$ for large cutoffs $N$ and $J$.}
The same holds for $\bracket{m}{k}$, taking $n\rightarrow m$, $j\rightarrow k$, and $\gb\rightarrow\qb$.  Note that once $\gb$ and $\qb$ have been chosen,\footnote{The choices of $\gb$ and $\qb$ can affect the numerical stability of the method, as will be discussed in future work.}
the calculation can be separated into a preprocessing step, in which the matrices $\bracket{n}{j}$, $\bracket{j}{n}$, and $\bracket{m}{k}$ are calculated (and potentially reused at subsequent steps of the cascade or for subsequent steps in an optimization), and the actual step of calculating $G^{jk}$ via Eqn.\ \ref{final}.

By exploiting the basis of the uncoupled system, we have reduced Eqn.\ \ref{mastern} to a set of simple linear algebraic equations,\footnote{More precisely, we have turned an $N^2 \times N^2$ matrix solve (where $N$ is a cutoff in copy count) into $K$ length-$J$ vector solves (where $J$ and $K$ are cutoffs in eigenmodes $j$ and $k$ respectively).} Eqn.\ \ref{final}, which dramatically speeds up the calculation without sacrificing accuracy (cf.\ {\it Results} and {\it Supplementary Material}).  The method is applicable for any input function $g_n$ and regulation function $q_n$.
Solutions using other bases and further generalizations to systems with feedback will be discussed in future work.

\section{Results}

\subsection{The spectral method is fast and accurate}

\ To demonstrate\footnote{All numerical procedures in the paper are implemented in MATLAB.} the accuracy and computational efficiency of the spectral method, we compare it both to an iterative numerical solution of Eqn.\ \ref{mastern} and to a stochastic simulation using the `Gillespie' algorithm \cite{Gillespie}
for a cascade of length $L=2$ with a Poisson input ($g_n=g=$ constant) and the discontinuous threshold regulation function
\beq
\label{thresh}
q_n=
\begin{cases}
q_- & {\rm for\ } n\le n_0 \\
q_+ & {\rm for\ } n>n_0.
\end{cases}
\eeq
The spectral method achieves an agreement up to machine precision with the iterative method in $\sim$$0.01 s$, which is $\sim$$1000$ times faster than the iterative method's runtime and $\sim$$10^8$ faster than the runtime necessary for the stochastic simulation to achieve the same accuracy; see {\it Supplementary Materials} for details.
The huge gain in computational efficiency over both the iterative method and the stochastic simulation makes the spectral method extremely useful, particularly for optimization problems, in which the probability function must be evaluated multiple times.
In the next sections we exploit this feature to optimize information transmission in signaling cascades.

\subsection{Information processing in signaling cascades}

\ Linear signaling cascades are a ubiquitous feature of biological networks,
used to transmit relevant information from one part of a cellular system to another \cite{Gomperts, Ting, Detwiler, Bolouri, Bassler}.  Information processing in a cascade is quantified by the mutual information \cite{Shannon}, which measures in bits how much information about an input signal is transmitted to the output signal in a noisy process. For a cascade of length $L$, the mutual information between an input species (with copy number $n_1$) and an output species (with copy number $n_L$)\footnote{Although the spectral method only involves the calculation of joint distributions between adjacent species in the cascade, the input-output distribution can be obtained using $p(n_1,n_\ell)=\sum_{n_{\ell-1}}p(n_1,n_{\ell-1})p(n_{\ell-1},n_\ell)/p(n_{\ell-1})$, initialized with $\ell=3$ and run up to $\ell=L$.  This assumes $p(n_1|n_{\ell-1},n_\ell)=p(n_1|n_{\ell-1})$, which at worst (at $\ell=3$) is equivalent to the Markovian approximation, Eqn.\ \ref{markov2}.}
is $I=\sum_{n_1,n_L} p(n_1,n_L) \log_2[p(n_1,n_L)/p(n_1)p(n_L)]$.
In this study we define the capacity $I^*$ as the maximum mutual information over either regulatory parameters, the input distribution, or both.  Depending on the signal to noise ratio, a high-capacity cascade functions either where the input signal is strongest or where the transmission process is least noisy \cite{Tkacik, Tkacik2}.

We first consider a cascade of length $L$ in which the regulation function $q_n$ is a simple threshold (Eqn.\ \ref{thresh}) with fixed parameters that are identical for each cascade step.
It is worth noting that while a threshold-regulated creation rate represents the simplest choice biologically, it is the most taxing choice computationally: as the discontinuity $\Delta=|q_+-q_-|$ increases, we find both that (a) a larger cutoff $K$ in eigenmodes is required for a desired accuracy, and (b) the accuracy of the Markovian approximation decreases (cf.\ {\it Supplementary Material}).  The results herein therefore constitute a stringent numerical challenge for the spectral method.

\begin{figure}
\begin{center}
\includegraphics[scale=0.45]{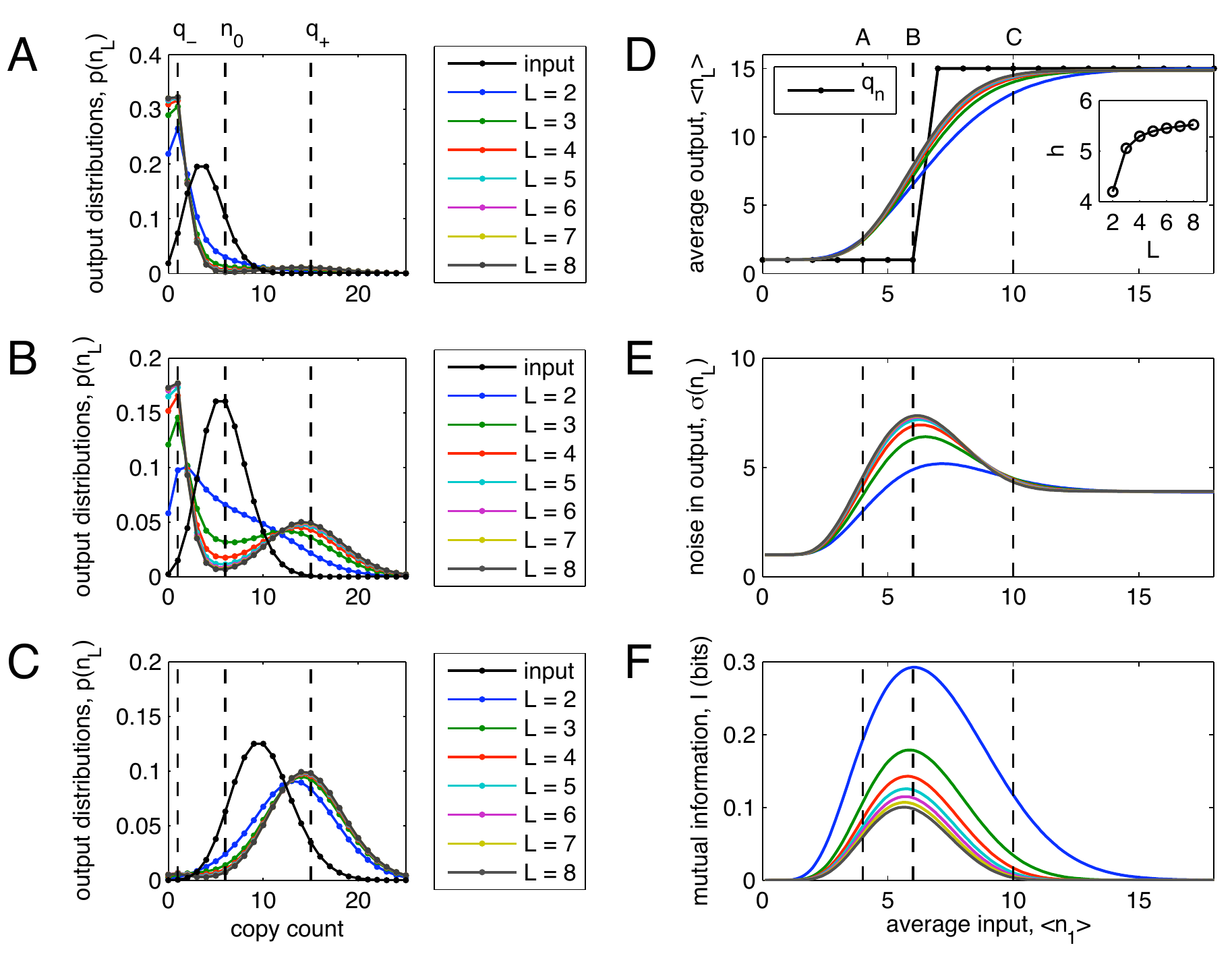}
\caption{Transfer functions and noise in a signaling cascade.  {\bf A-C:} Plots of input distribution $p(n_1)$ (black) and output distributions $p(n_L)$ (colors; see legend) for various cascade lengths $L$.  Input distribution is a Poisson centered at $\av{n_1}=4$ (in A), $\av{n_1}=n_0=6$ (in B), or $\av{n_1}=10$ (in C).  The regulation function $q_n$ for all steps is a threshold (Eqn.\ \ref{thresh}) shown in D (black line with dots), with parameters $q_-$, $q_+$, and $n_0$ overlaid as dashed lines in A-C.  The degradation rate ratio is $\rho=1$, and Eqn.\ \ref{mastern} is solved using the spectral method with $\qb=10$ and $\gb=\langle g_n\rangle$ for each step in the cascade.  {\bf D-F:} Transfer functions (average output $\av{n_L}$) (D), noise (standard deviation of the output $\sigma(n_L)$) (E), and mutual information $I$ (F) as functions of average input $\av{n_1}$ for various cascade lengths $L$ (colors as in A-C).  As in A-C, input is Poisson at every $\av{n_1}$; dashed lines correspond to the specific $\av{n_1}$ values in A, B, and C.  {\bf Inset in D:} Cooperativity $h$ as a function of $L$.
}
\label{transfer}
\end{center}
\end{figure}

We take the input $p(n_1)$ to be a Poisson distribution (i.e.\ $g_n=g=$ constant).
In the extreme cases, when the threshold is infinite or zero, the output is a Poisson distribution centered at $q_-$ or $q_+$, respectively.  Similarly, when the input median is below or above threshold, the output mean should be biased toward $q_-$ or $q_+$, respectively.  For example, in Fig.\ \ref{transfer}A, $\av{n_1}<n_0$, and the output distribution is shifted toward $q_-$.  This effect is amplified at each step of the cascade, such that $\av{n_L}\to q_-$ for large $L$.  Similarly,  $\av{n_L}\to q_+$ for large $L$ when $\av{n_1}>n_0$
(Fig.\ \ref{transfer}C).  When $\av{n_1}\sim n_0$ (Fig.\ \ref{transfer}B), the output is balanced between $q_-$ and $q_+$; if the discontinuity $\Delta=|q_+-q_-|$ is sufficiently large, the output is bimodal, as discussed in more detail in the next section.

Mutual information $I$ decreases monotonically with $L$ for all $\av{n_1}$ (cf.\ Fig.\ \ref{transfer}F), as required by the data processing inequality \cite{Cover} (i.e.\ one cannot learn more information from the output of an ($L$+1)-gene cascade than one could from an $L$-gene cascade with identical regulation, only less).  $I$ is maximal for $\av{n_1}\sim n_0$ which makes intuitive sense, as it corresponds to the input taking advantage of both rates $q_-$ and $q_+$ roughly equally in producing the output. A simple calculation quantifies this intuition.  Approximating the steady-state distribution for the moment as a strict switch conditional on $n_0$ (i.e. $p(n_L|n_1)=p_-(n_L)$ if $n_1<n_0$ and $p(n_L|n_1)=p_+(n_L)$ if $n_1\ge n_0$ for some distributions $p_\pm(n_L)$), it follows from the definition of $I$ that
\beq
\label{switch}
I_{\rm switch}=S-\sum_{\pm} \sum_{n_L} p_{\pm}(n_L)
	\log_2\left[1+\frac{\pi_\mp p_{\mp}(n_L)}{\pi_\pm p_{\pm}(n_L)}\right],
\eeq
where $S=-\sum_\pm\pi_\pm\log_2\pi_\pm$, $\pi_-=\sum_{n_1<n_0}p(n_1)$ and $\pi_+=\sum_{n_1\ge n_0}p(n_1)=1-\pi_-$.  If there is little overlap between $p_-(n_L)$ and $p_+(n_L)$, then $I_{\rm switch}\sim S$, which is maximal when $\pi_-=\pi_+$, i.e.\ when the median of the input distribution $p(n_1)$ lies at the threshold $n_0$.  Additionally, since the maximal value of $S$ (and $I_{\rm switch}$, since the summand of the second term in Eqn.\ \ref{switch} is always nonnegative) is $1$ bit, this calculation also suggests that the capacity of threshold regulation (in the limit of strict switch-like behavior) is limited to $1$ bit.  Again, this result is intuitive, as the cascade is only passing on the binary information of whether particles in the input distribution are below or above the threshold.

In an experimental setup one might have access to only the mean response (or ``transfer function''), or the variance in response across cells (or ``noise''), of a signaling cascade to its input.  Since our method yields the full distributions, such summary statistics are readily computed.  
Despite the sharpness of threshold regulation, the transfer functions are quite smooth even at $L=2$ (cf.\ Fig.\ \ref{transfer}D).  The effect of the intrinsic noise
is to smooth out a sharp discontinuity in creation rates, producing a continuous mean response. 
The transfer functions shown are least-squares fit to Hill functions of the form $\bracket{n_L}{n_1}=\alpha_-+(\alpha_+-\alpha_-)(n_1)^h/[(n_1)^h+(k_d)^h]$.
As one would expect, for all $L$, best fit values of $\alpha_-$, $\alpha_+$ are near the rates $q_-$ and $q_+$ respectively, and best fit $k_d$ values are near the threshold $n_0$.  As $L$ increases, the transfer function sharpens, and cooperativity $h$ increases (cf.\ inset in D), due to the amplified migration of the output to either $q_-$ and $q_+$ in longer cascades (as in Figs.\ \ref{transfer}A and \ref{transfer}C).

The strength of the noise increases with $L$ (cf.\ Fig.\ \ref{transfer}E), consistent with the reduction in $I$ with $L$ (cf.\ Fig.\ \ref{transfer}F),
and the noise is peaked at the threshold.
The ``switch approximation" (cf. Eqn.\ \ref{switch}) illustrates the gain in information
when the median of the input coincides with the threshold; the ``small noise approximation" \cite{Tkacik,Tkacik2}, however, illustrates the loss in information when the peak of the input coincides with the peak of the noise. The tradeoff between these two trends thwarts information transmission with unimodal input distributions (e.g., those used in Fig.\ \ref{transfer}) and
suggests an input distribution with two or more modes should be able to transduce
more information. Such distributions are the subject of study below,
and, in related work \cite{Gregor, Tkacik2} are shown to be the optimal strategy and to be
observed in biology for a regulatory system in which peak noise and threshold coincide.

\subsection{Bimodal output from a unimodal input}

\ A striking feature of Fig.\ \ref{transfer}B is that the unimodal input is converted to a bimodal output for cascades of length $L=3$ or longer.
Bimodality can arise from a system with two genes whose proteins repress each other or from a single gene whose proteins activate its own expression.
Here we demonstrate that cascades with sufficiently strong regulation constitute an information-optimal mechanism for a cell to achieve bimodality.

\begin{figure}
\begin{center}
\includegraphics[scale=0.45]{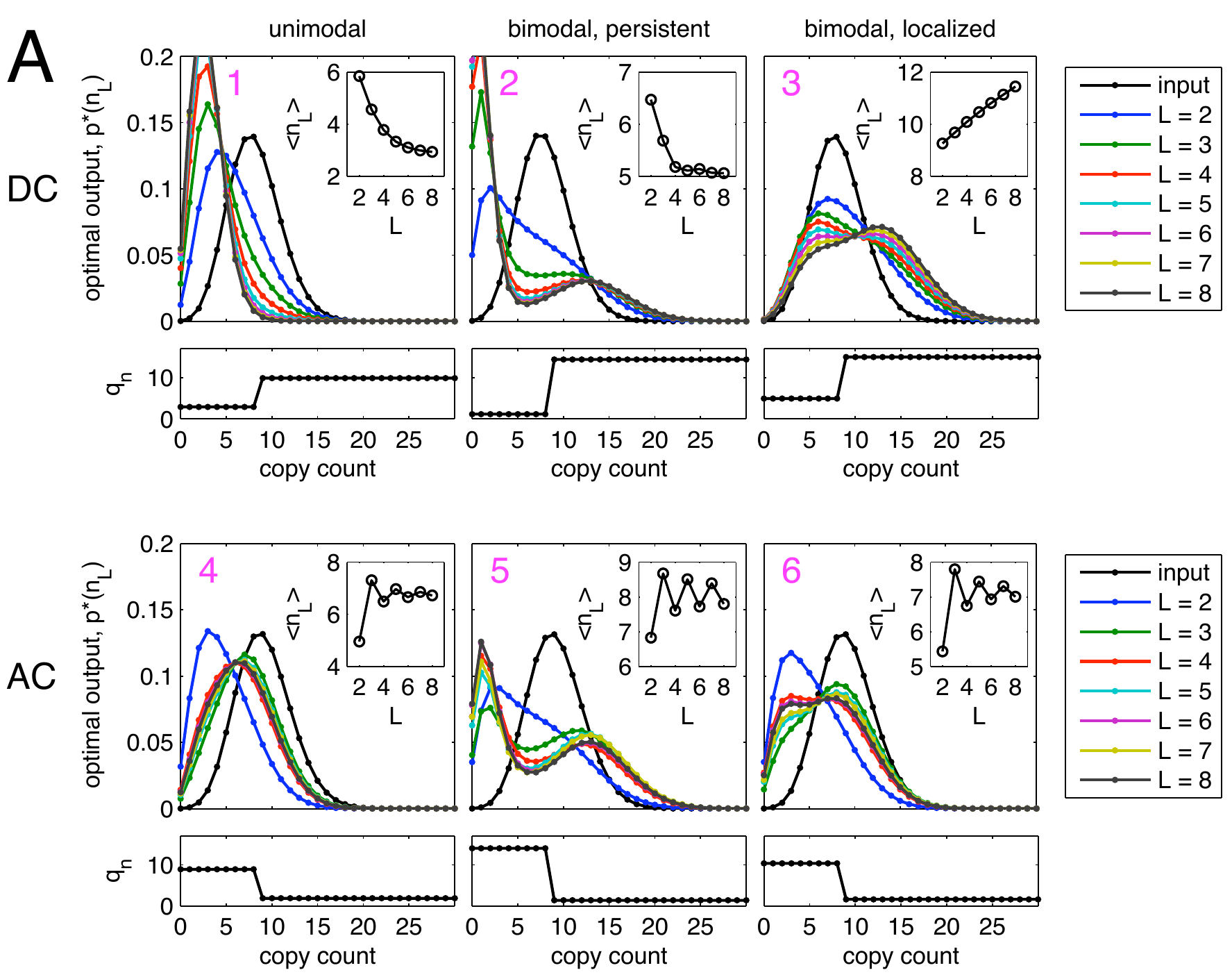} $\quad$
\includegraphics[scale=0.48]{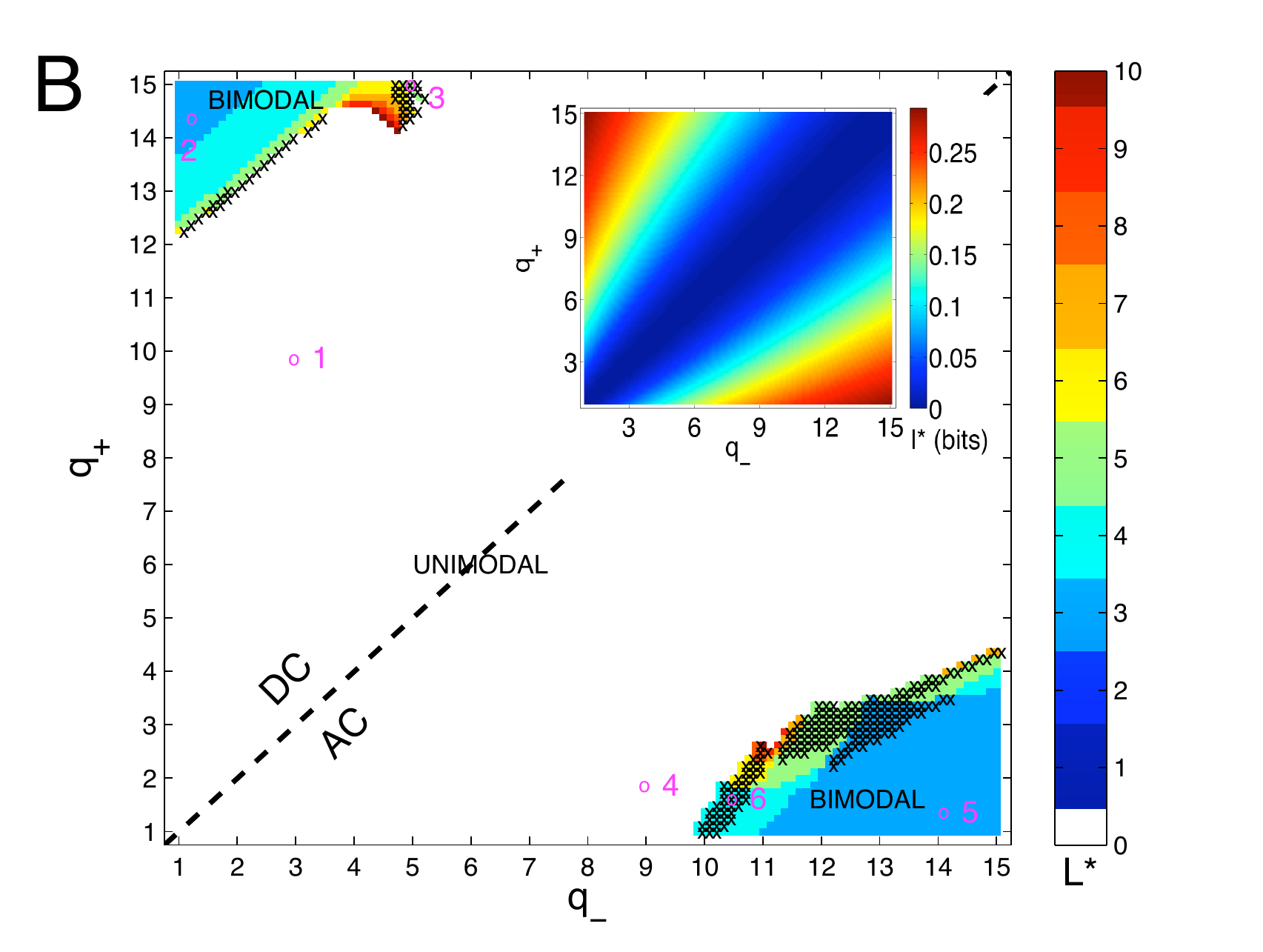}
\caption{Optimal output modality in cascades with unimodal input.  {\bf A:} Plots of optimal input distribution $p^*(n_1)$ (black) and optimal output distributions $p^*(n_L)$ (colors; see legend) for different cascade lengths $L$ (optimal input distributions are qualitatively the same; only that for $L=2$ is shown), corresponding to regulation functions $q_n$ (identical for each step) plotted underneath ($\rho=1$, and solutions used $\qb=10$ and $\gb=\langle g_n\rangle$).  Mutual information $I$
is optimized as a function of the mean $g$ of the Poisson input distribution.  Magenta numbers on plots correspond to magenta points in B.  Insets show plots of average output $\langle n_L \rangle$ vs.\ cascade length $L$.  In the first column of A, the output is always unimodal; in the second column, the output is bimodal for cascade lengths $L\ge L^*$ for some $L^*$ (``persistent'' bimodality); in the third column, the output is bimodal for a range of $L$ values, then unimodal once more for large $L$ (``localized'' bimodality).  The first row shows ``DC'' cascades, in which each step is up-regulating, and the second row shows ``AC'' cascades, in which each step is down-regulating.  {\bf B:} Phase diagram of optimal output modality as a function of $q_-$ and $q_+$
($n_0=8$).  White is unimodal, and color is bimodal, with color corresponding to $L^*$.
Distinction between persistent (no `x') and localized (`x') bimodality is shown up to $L=10$.  Dashed line separates DC cascades from AC cascades.
{\bf Inset:} Capacity $I^*$ in bits as a function of $q_-$ and $q_+$ for the same data.}
\label{bimod}
\end{center}
\end{figure}

Recall that Fig.\ \ref{transfer}B corresponds to the case where the input distribution is optimally matched with the regulation function, i.e.\ the bimodal output represents optimal information transmission.  By optimizing over the mean $g$ of a Poisson input distribution, we find that the most informative output distribution in a cascade with unimodal input can be unimodal or bimodal, depending on regulatory parameters and the length of the cascade.  Fig.\ \ref{bimod}A shows examples of regulation functions which produce output distributions that are unimodal, bimodal for cascades as long or longer than some $L^*$ (which we term ``persistent'' bimodality), and bimodal for short cascades but unimodal at both initial and final nodes for longer cascades (which we term ``localized'' bimodality).  

Bimodality is found both in cascades in which each step is down-regulating, which we call ``AC'' cascades, and in those in which each step is up-regulating, which we call ``DC'' cascades.  In DC cascades, as seen in the insets of panels 1-3 in Fig.\ \ref{bimod}A, the average output either monotonically decreases or monotonically increases with $L$.  In the former case, since $q_-<n_0$, the probability that the output is below the threshold given that the input is below threshold is large. Successive such regulations drive the probability of being below the threshold towards $1$, successively decreasing $\av{n}$ at each step in the cascade.  In the latter case, since $q_+>n_0$, the same picture holds, and $\av{n}$ monotonically increases with $L$.  Whether the monotonically increasing or decreasing behavior is the more informative is determined by the relationship among $q_+$, $q_-$, and $n_0$.
In AC cascades, an analogous picture holds but with alternation: $\pi_-<\pi_+$ for the even-numbered links (cf.\ Eqn.\ \ref{switch}), and the AC condition $q_->q_+$ leads to $\pi_+<\pi_-$ for the odd-numbered links, as illustrated in the insets of panels 4-6 in Fig \ref{bimod}A.  These behaviors motivate the names ``AC'' and ``DC,'' analogous to alternating and direct current flow.  Performance of AC and DC cascades is compared in more detail in the next section.

Fig.\ \ref{bimod}B shows a phase diagram of optimal output modality as a function of the rates $q_-$ and $q_+$: bimodality is found at high values of the discontinuity $\Delta=|q_+-q_-|$ (specifically, for $\Delta\gtrsim11$ in AC circuits and $\Delta\gtrsim12$ in DC circuits when $n_0=8$).  Intuitively, since the weight of the output is distributed between $q_-$ and $q_+$ for long cascades, increasing their separation spreads the weights apart and creates bimodal distributions.  Furthermore, as $\Delta$ increases, the bimodality becomes more robust: it goes from localized to persistent, and its onset occurs at a smaller cascade length $L^*$.  The inset of Fig.\ \ref{bimod}B shows that capacity $I^*$ also increases with $\Delta$; cascades with bimodal output therefore have higher capacities than those with unimodal output.  As $\Delta$ increases, the information transmission properties of a regulatory cascade are better approximated by simple switch-like regulation (cf.\ Eqn.\ \ref{switch}). In short, summarizing the input distribution by $\pi_+$ and $\pi_-$
is a more informative summary of the distribution as the regulation becomes more discontinuous.

\subsection{Channel capacity in AC/DC cascades}

\ Our setup provides a way to ask quantitatively whether a cascade with down-regulating steps (AC) can transmit information with more or less fidelity than a cascade with down-regulating steps (DC).  Since a cell must expend time and energy to make proteins, a fair comparison between cascade types can only be made when the species involved in each type are present in equal copy number.  As in \cite{Ziv}, we introduce the objective function $\L = I - \lambda\langle n\rangle$
where $I$ is mutual information
and $\langle n\rangle=\sum_{\ell=1}^L\langle n_\ell\rangle/L$ is an average copy count over all species in the cascade.  Here $\lambda$ represents the metabolic cost of making proteins, and optimizing $L$ for different values of $\lambda$ allows a comparison of AC and DC capacities $I^*$ at similar values of $\langle n\rangle$.

For both AC and DC cascades, $I^*$ increases with $\n$ as more proteins are made available to encode the signal,\footnote{There is a slight decrease in $I^*$ with $\n$ beginning near $\n\approx8$ that is more pronounced at higher $L$.  This is likely due to the decrease in accuracy of the Markovian approximation with increasing $\Delta$ (cf.\ {\it Supplementary Material}), since large $\n$ requires large $\Delta$.
Calculations with the full joint distribution (via stochastic simulation)
at $L=3$ and $4$ give qualitatively similar results, but with $I^*$ increasing monotonically with $\n$.}
and $I^*$ decreases\footnote{The decrease of $I^*$ with $L$ is consistent with, but not a direct consequence of, the data processing inequality \cite{Cover}, as each $p^*(n_1,\dots,n_L)$ results from a separate optimization for each subsequent choice of $L$.} with $L$ at all $\langle n\rangle$
(cf.\ Fig.\ \ref{acdcZ}A).
Both AC and DC capacities converge to an $L$-dependent asymptotic value at high copy count, but DC cascades attain higher capacities per output protein than AC cascades.  The difference is most pronounced at low copy count ($\n\lesssim7$), and more pronounced still for longer cascades.
The difference is easily explained: AC and DC cascades of the same length with the same discontinuity $\Delta=|q_+-q_-|$ have the same capacity but have different mean numbers of proteins. Recall from Fig.\ \ref{bimod}B that large $\Delta$ leads to
high-capacity,
bimodal solutions. The difference between AC and DC cascades is in the placement of their optimal distributions for a given $\Delta$.  We observe that optimal
AC cascades tend to exhibit $\av{n}\gtrsim n_0$, while optimal DC cascades tend to exhibit $\av{n}\lesssim n_0$.  Ultimately, this allows DC cascades to achieve the same capacity for the same regulation parameters (cf.\ Fig.\ \ref{bimod}B inset), but use fewer proteins.  These results suggest that DC cascades  transmit with higher fidelity per protein than AC cascades when protein production is costly.

\subsection{The most informative input to a threshold is bimodal}

\ If the first species is governed by more than a simple birth-death process, the input to a cascade will not be a simple Poisson distribution.
To investigate the role of input multimodality in information transmission, we consider inputs defined by a mixture of Poisson distributions, $p(n_1) = \sum_{i=1}^Z \pi_i e^{-g_i}g_i^{n_1}/n_1!$,
(with $\sum_{i=1}^Z \pi_i=1$).
As before, we expect information to increase with copy number, and we use the objective function $\L$ 
when optimizing over the input distribution $p(n_1)$.

\begin{figure}
\begin{center}
\includegraphics[scale=0.45]{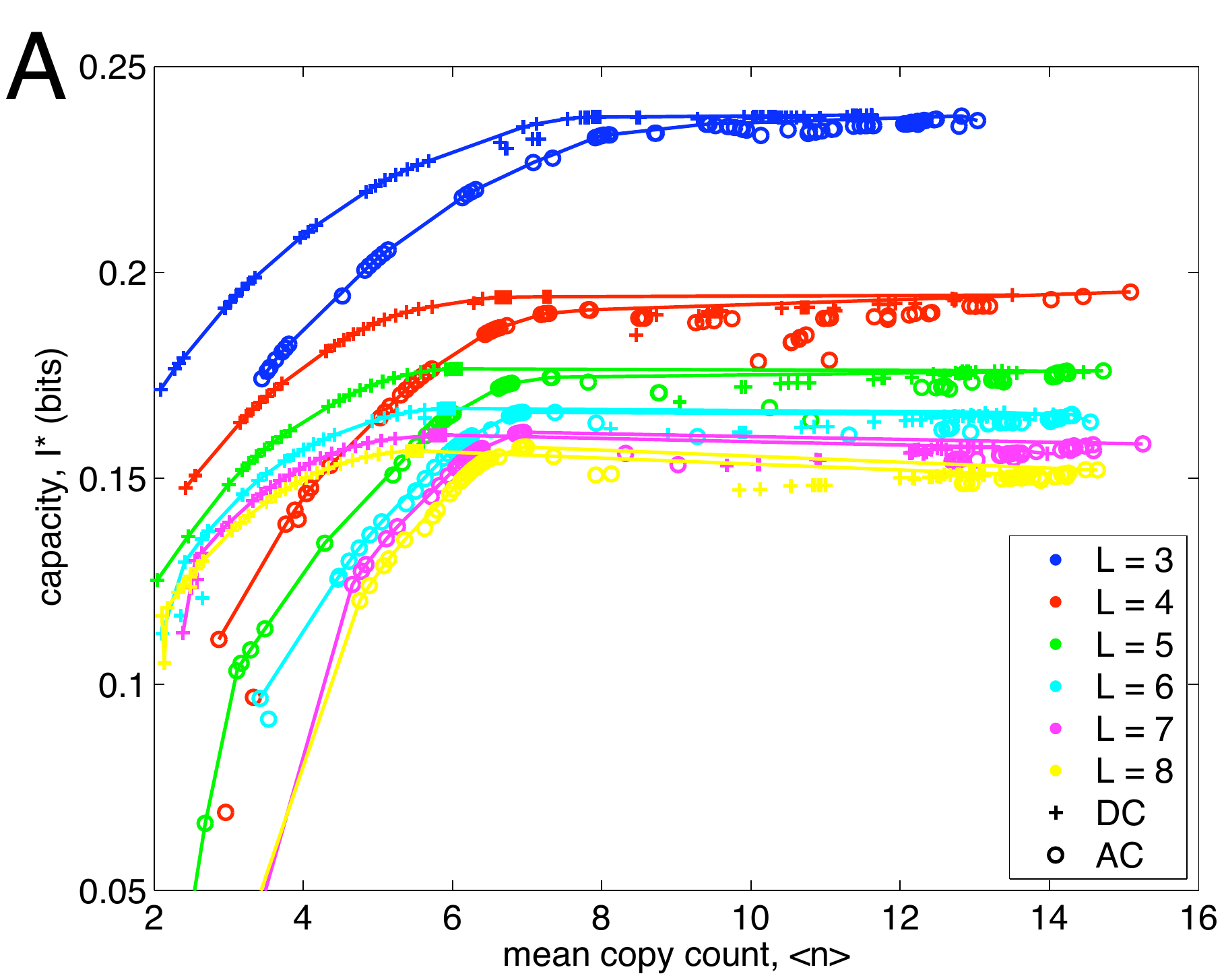}
\includegraphics[scale=0.45]{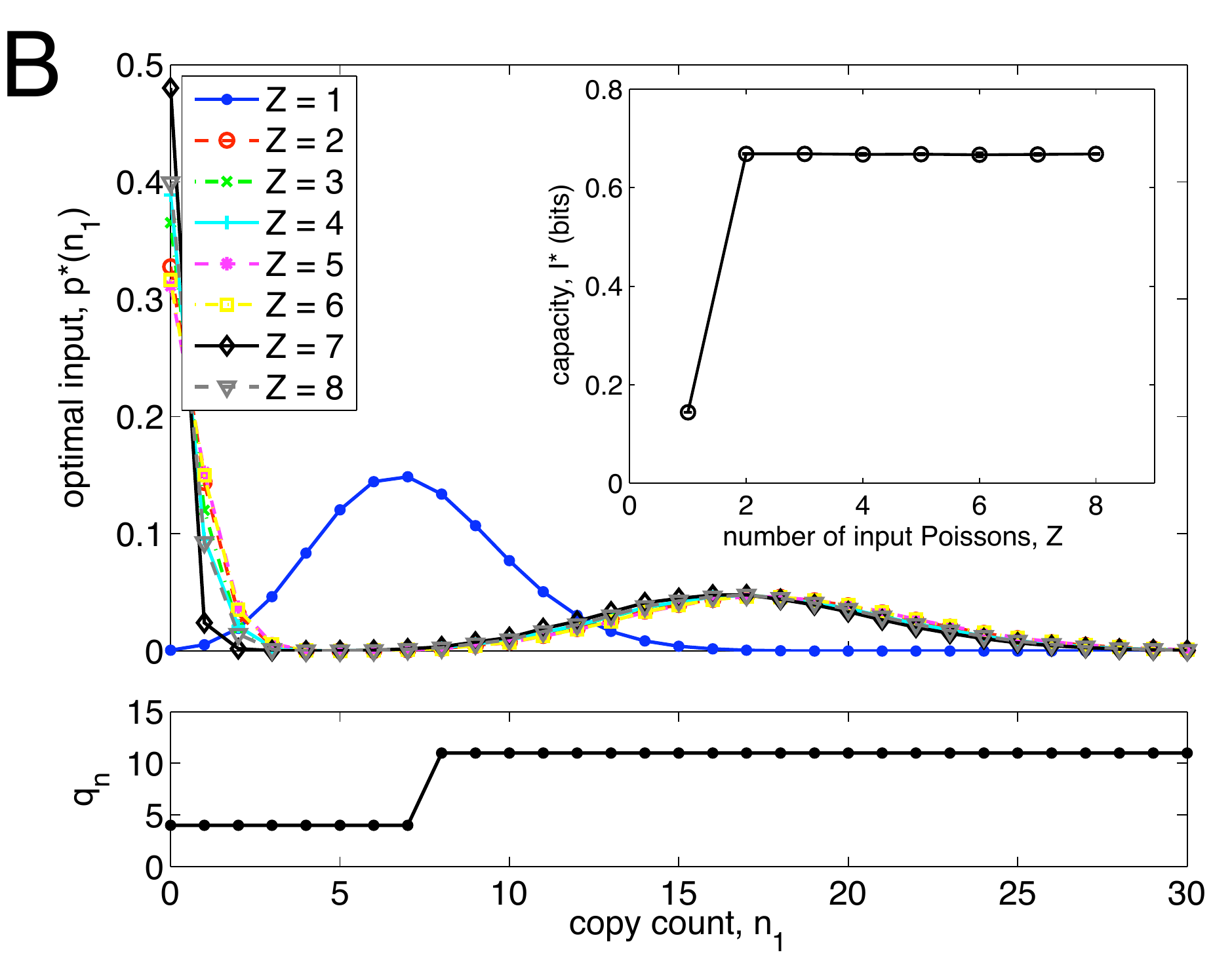}
\caption{{\bf A:} Capacities of AC and DC cascades.  Capacity $I^*$ vs.\ average copy count $\langle n \rangle$ (over all species) for AC (circles) and DC (plus signs) cascades of different lengths $L$ (color), with Poisson input distributions.  Results were obtained by optimizing the objective function $\L$ 
over all parameters ($q_-$, $q_+$, $n_0$, and $g$) for $\lambda=1\times 10^{-6}-3\times 10^{-2}$ ($n_0$ is constrained to be an integer, the regulation function is the same for every step, $\rho=1$, and solutions used $\qb=10$ and $\gb=\langle g_n\rangle$).  Lines show convex hulls. {\bf B:} Optimal input distributions with different numbers $Z$ of Poisson distributions.  The cascade length is $L=2$, the degradation rate ratio is $\rho=1$, and the regulation function $q_n$ is plotted in the bottom panel; solutions used $\qb=10$ and $\gb=\langle g_n\rangle$.  The objective function $\L$ 
is optimized with $\lambda=10^{-4}$ over all input parameters $g_i$ and $\pi_i$.  {\bf Inset} Capacity $I^*$ vs.\ $Z$, averaged at each $Z$ over 7 optimizations with different initial conditions.
}
\label{acdcZ}
\end{center}
\end{figure}

All optimal input distributions with $Z\ge2$ are bimodal, with one mode on either side of the threshold (cf.\ Fig.\ \ref{acdcZ}B).  When $Z$ is $3$ or more, either all but two $\pi_i$ values are driven by the optimization to $0$, or all the $g_i$ values with nonzero weights are driven to one of two unique values.  The two modes are roughly equally weighted (i.e.\ $\pi_1\approx\pi_2\approx0.5$), consistent once more with our calculation that the median of the optimal input distribution falls roughly at the threshold (cf.\ Eqn.\ \ref{switch}).  As an additional verification, a plot of capacity $I^*$ vs.\ $Z$ in the inset of Fig.\ \ref{acdcZ}B reveals that $I^*$ remains constant for $Z=2$ and beyond.  A threshold regulation function presents a binary choice, and the optimal input is a bimodal distribution that equally utilizes both sides.
Lastly, we point out that the capacities in the inset of Fig.\ \ref{acdcZ}B are below $1$ bit.  Even with a bimodal input distribution, a short cascade ($L=2$), and strong regulation functions (i.e.\ large discontinuities $\Delta$), we do not find capacities above $1$ bit.  This is consistent with our calculation under the approximation of threshold regulation as a switch (cf.\ Eqn.\ \ref{switch}) and with the intuition that a threshold represents a binary decision.

\section{Conclusions}

We have introduced a method, the {\it spectral} method, 
which exploits the linear algebraic structure of the master equation,
and expands the full problem in
terms of its natural eigenfunctions.  
We have illustrated our method by probing the optimal transmission
properties of
signaling cascades with threshold regulation.  We have shown that sufficiently
long cascades with sufficiently strong regulation functions optimally convert a unimodal input to a
bimodal output.  
A bimodal input is optimal for
information transmission across a threshold, 
and a multimodal
input offers no further processing power. Sustained bimodality of the
input distribution requires large discontinuities $\Delta$ between the production
rates below and above the threshold. The value of $\Delta$
controls the maximum information transmitted by a cascade with
threshold regulation in a similar way for cascades of up-regulations (DC)
and cascades of down-regulations (AC), but a DC cascade
outperforms an AC cascade by 
using fewer average copies of its species.
We emphasize that the application of the spectral method to signaling
cascades represents only a beginning.  Variations on the natural bases
in which to expand, and extensions of the method to other small
network topologies, will be the subject of future work.  More
generally, however, we anticipate that the method will prove useful in
the direct solution of a large class of master equations describing a
wide variety of biological systems.

\section{Supplementary Material}

\subsection{The spectral method is fast and accurate}

\begin{figure}
\begin{center}
\includegraphics[scale=0.4]{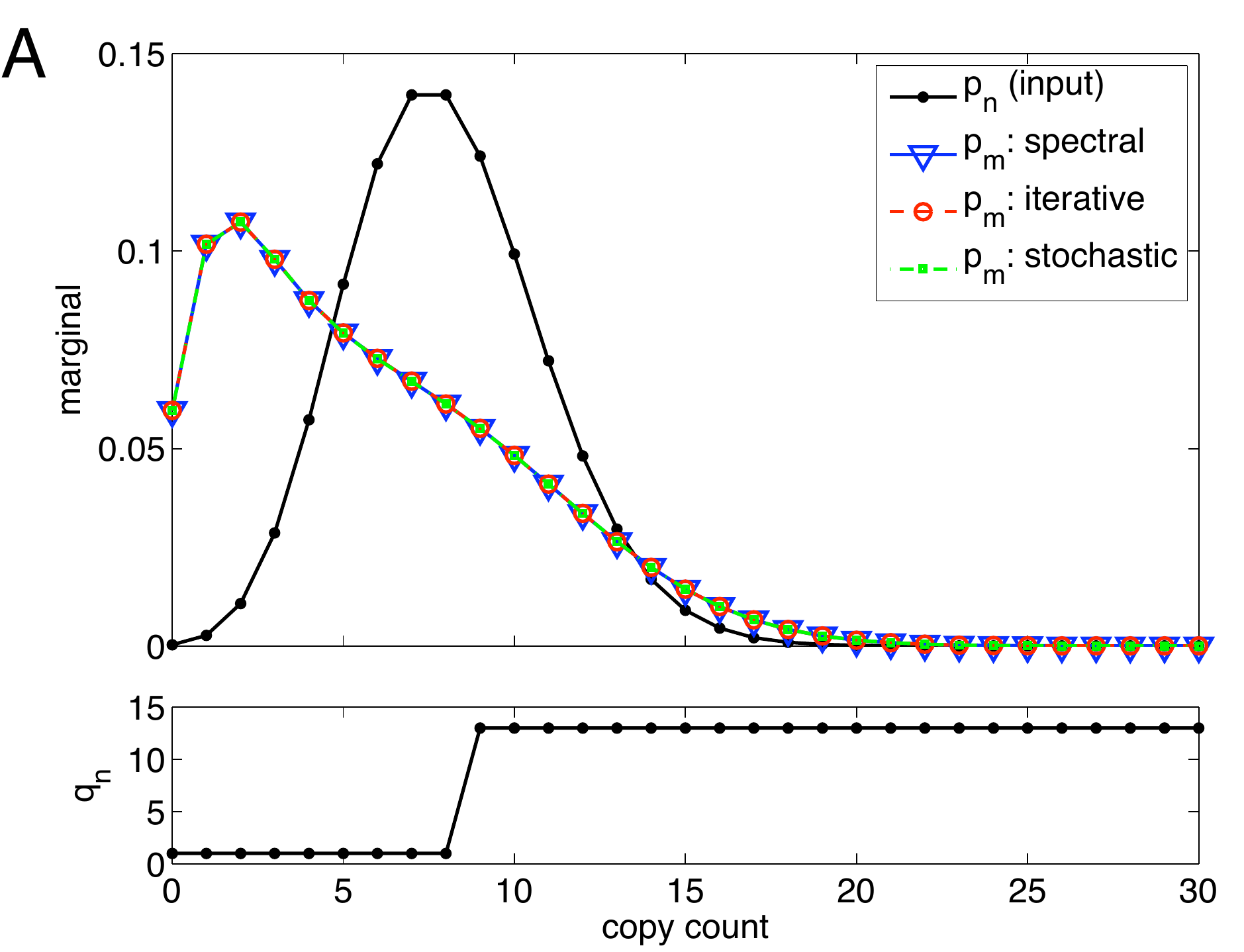} $\quad$
\includegraphics[scale=0.4]{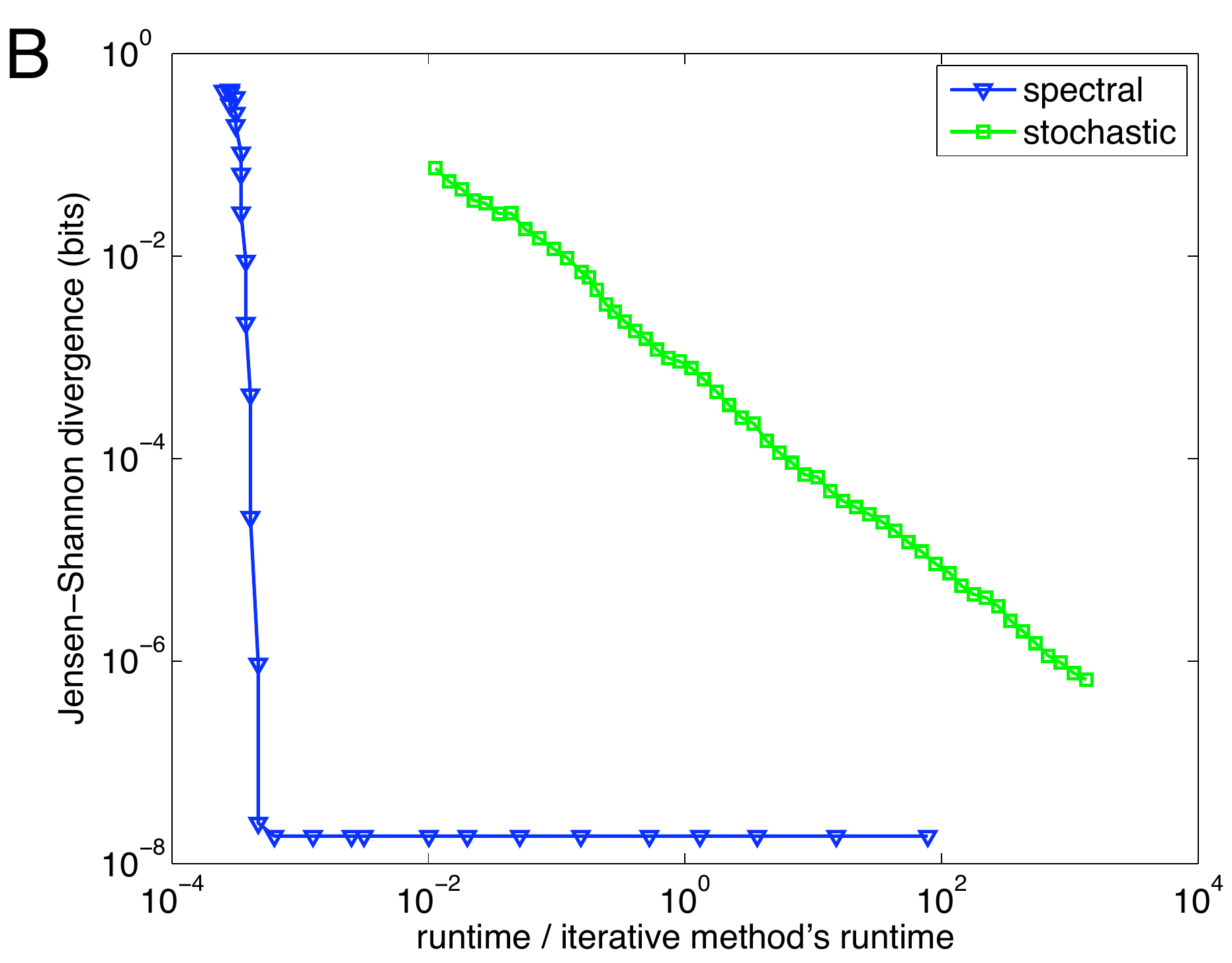}
\linespread{1}
\caption{Demonstration of the spectral method's accuracy and efficiency.  {\bf A:} Output distributions $p_m$ obtained by solving Eqn.\ \ref{mastern} with $g_n=g=$ constant using the spectral method described in the main text (blue triangles), an iterative numerical method (red circles), and a stochastic simulation \cite{Gillespie} (green squares).  Also shown is the input $p_n$ (a Poisson distribution with mean $g=8$), and the regulation function $q_n$ (Eqn.\ \ref{thresh}) with $q_-=1$, $q_+=13$, and $n_0=8$; the degradation rate ratio is $\rho=1$. Spectral solution used $\bar{q}=10$ and mode number cutoffs $J=K=50$, and all solutions used copy count cutoffs $N=M=50$.  {\bf B:} Jensen-Shannon divergence \cite{Lin} between $p_{nm}$ obtained using the iterative numerical method and that obtained using the spectral method (blue triangles) or stochastic simulation (green squares), plotted against the latter two methods' respective runtimes.  Runtimes are scaled by iterative method's runtime, $15.9$ s.  Spectral method data obtained by varying $K$ from $3$ to $12,589$; plateau begins at $K\approx50$.  Stochastic simulation data obtained by varying integration time from $100$ to $2\times10^7$, in units of the upstream gene's reciprocal degradation rate.}
\label{check}
\end{center}
\end{figure}

To demonstrate the accuracy and computational efficiency of the spectral method, we compare it both to an iterative numerical solution and to a stochastic simulation (using the varying step Monte Carlo or `Gillespie' method \cite{Bortz, Gillespie}) of Eqn.\ \ref{mastern}.
Fig.\ \ref{check}A shows the agreement among output distributions $p_m$ generated by the three methods for a cascade of length $L=2$ with a Poisson input ($g_n=g=$ constant) and the threshold regulation function in Eqn.\ \ref{thresh}. The spectral method achieves accuracy up to machine precision in $\sim$0.01 s, which is $\sim$$1000$ times faster than the iterative method's runtime and $\sim$$10^8$ faster than the runtime necessary for the stochastic simulation to achieve the same accuracy (cf.\ Fig.\ \ref{check}B).  
As a measure of error we use the Jensen-Shannon divergence \cite{Lin} (a measure in bits between two probability distributions) between the distribution $p_{nm}$ generated by the iterative method and that generated by either the spectral method or the stochastic simulation.  We plot this measure against the runtime of either method, scaled by the runtime of the iterative method, in Fig.\ \ref{check}B.
 
\subsection{Validity of the Markovian approximation}

For a cascade of length $L$, we reduce an $L$-dimensional master equation to a set of 2-dimensional equations by employing the Markov approximation, i.e.\ that each species is conditionally independent of distant nodes given proximal nodes (cf.\ main text).  Here we compare results under this approximation with those from a solution of the full master equation, using both a stochastic simulation \cite{Gillespie} and a non-Markovian implementation of the spectral method.

The non-Markovian spectral method is implemented as follows.  The full master equation for the process $n_1\xrightarrow{q_2(n_1)}n_2\xrightarrow{q_3(n_2)}\dots\xrightarrow{q_L(n_{L-1})}n_L$ is
\beqn
\label{masterL}
\dot{p}(\vec{n}) &=&gp(\vec{n}-\hat{e}_1)-gp(\vec{n})+(n_1+1)p(\vec{n}+\hat{e}_1)-n_1p(\vec{n})\nonumber\\
&&+\sum_{\ell=2}^L\rho_\ell\left[q_\ell(n_{\ell-1})p(\vec{n}-\hat{e}_\ell)-q_\ell(n_{\ell-1})p(\vec{n})\right.\nonumber\\
&&\left.+(n_\ell+1)p(\vec{n}+\hat{e}_\ell)-n_\ell p(\vec{n})\right],
\eeqn
where time is rescaled by the degradation rate of the first species, $g$ is the creation rate of the first species, $q_\ell$ is the creation rate of the $\ell$th species, creation rates are normalized by corresponding degradation rates, $\rho_\ell$ is the ratio of the degradation rate of the $\ell$th species to that of the first, $\vec{n}=(n_1,\dots,n_L)$, and $\hat{e}_\ell$ represents a $1$ in $n_\ell$th direction.  
Denoting $\ket{j_1,\dots,j_L}$ as the eigenstate of species $\vec{n}$ at constant rates $g,\qb_2,\dots,\qb_L$ respectively, Eqn.\ (\ref{masterL}) has the spectral decomposition
\beqn
\label{GL}
0&=&\left(j_1+\sum_{\ell=2}^L\rho_\ell j_\ell\right)G^{j_1,\dots,j_L}\nonumber\\
	&&+\sum_{\ell=2}^L\rho_\ell\sum_{j'_{\ell-1}}\Delta_{j_{\ell-1},j'_{\ell-1}}
	G^{j_1,\dots,j'_{\ell-1},j_\ell-1,\dots,j_L},
\eeqn
where $G^{j_1,\dots,j_L}=\bracket{j_1,\dots,j_L}{G}$ and the deviation operator $\Delta_{j_{\ell-1},j'_{\ell-1}}=\sum_{n_{\ell-1}} \left[\qb_\ell-q_\ell(n_{\ell-1})\right]\bracket{j_{\ell-1}}{n_{\ell-1}}\bracket{n_{\ell-1}}{j'_{\ell-1}}$.  Eqn.\ (\ref{GL}) is solved by iteration, initialized with $G^{j_1,\dots,j_L}=\delta_{j_1,0}\dots\delta_{j_L,0}$.  The joint distribution is obtained via inverse transform:
\beq
p(n_1,\dots,n_L)=\sum_{j_1,\dots,j_L} \bracket{n_1}{j_1}\dots\bracket{n_L}{j_L} G^{j_1,\dots,j_L}.
\eeq

\begin{figure}
\includegraphics[scale=0.41]{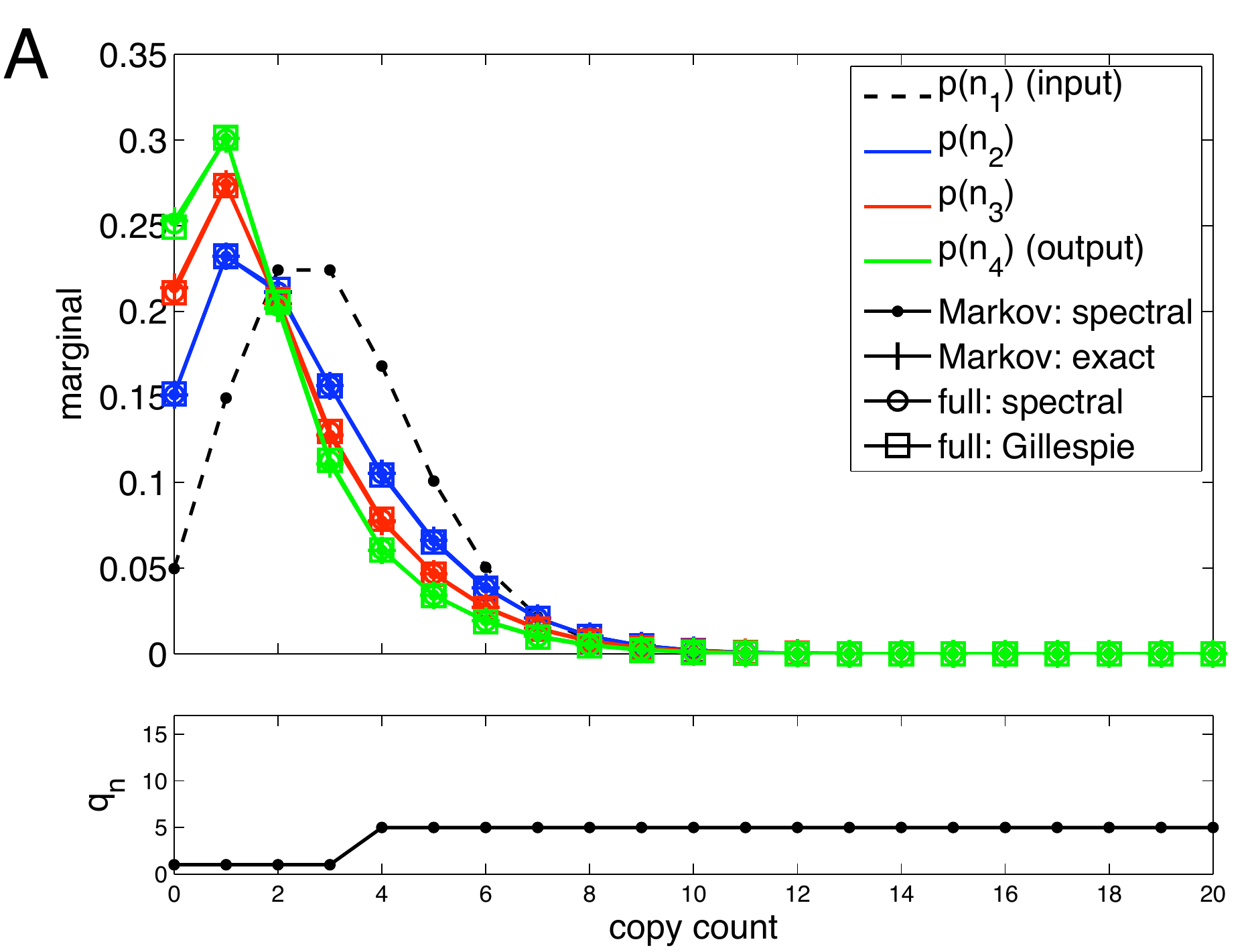} $\quad$
\includegraphics[scale=0.41]{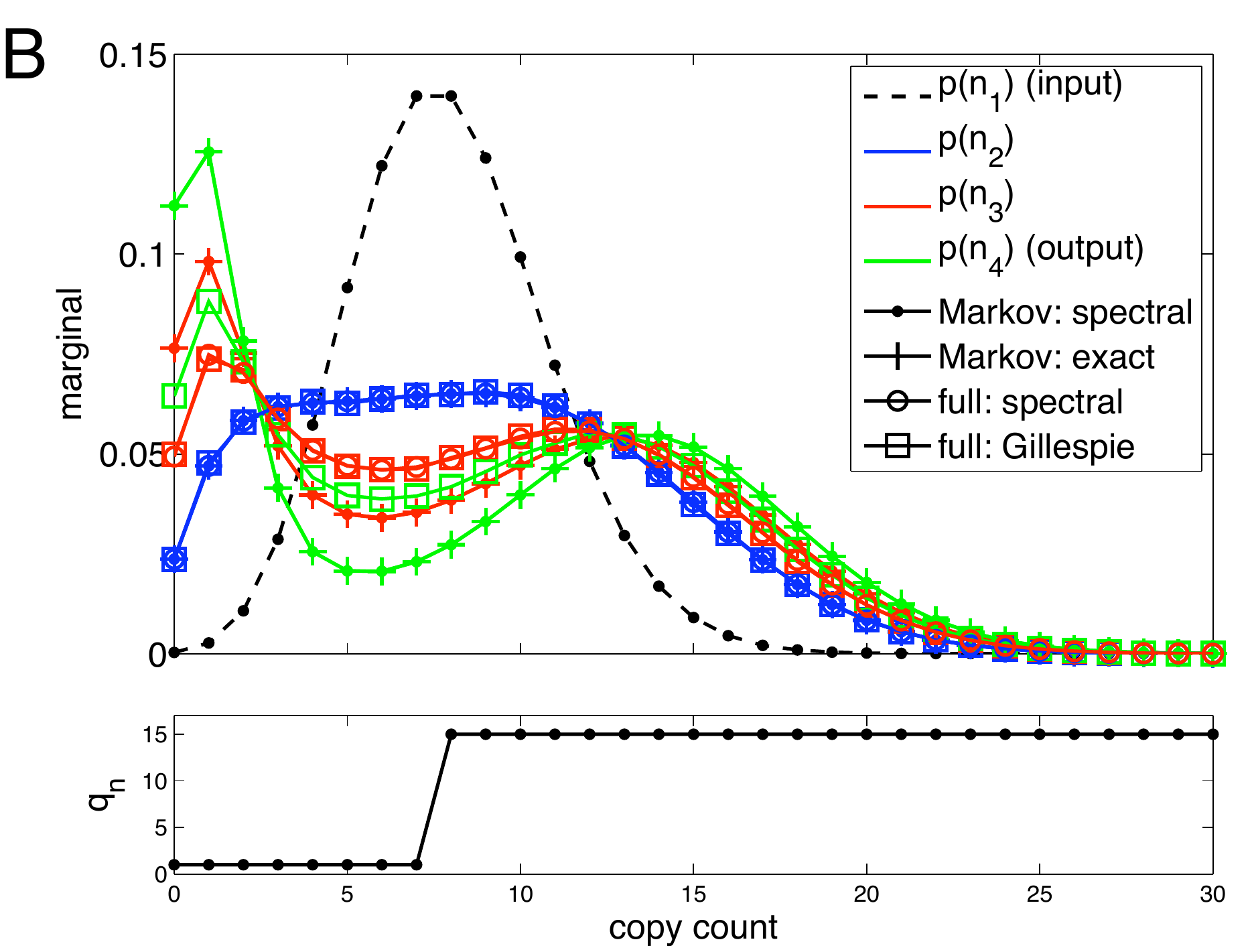}
\linespread{1}
\caption{Marginal distributions from a Markovian and a non-Markovian stochastic description of a cascade.  {\bf A:} Marginal distributions for the species in a cascade of length $L=4$ (colors indicate order of species in the cascade; see legend), with threshold regulation function $q_n$ shown in the bottom panel ($q_-=0.5$, $q_+=5$, and $n_0=7$).  Distributions were obtained using the Markov approximation (cf.\ main text), either via the spectral method (dots) or via an iterative numerical solution (plus signs), and using the full master equation (Eqn.\ \ref{masterL}), either via the non-Markovian spectral method (circles) or via stochastic simulation \cite{Gillespie} (squares).  {\bf B:} As A, but with $q_+=9$.}
\label{markov}
\end{figure}

For a cascade of length $L=4$, with two different values of the discontinuity $\Delta=|q_+-q_-|$ (cf.\ Eqn.\ \ref{thresh}), Figure \ref{markov} compares the marginal distributions calculated under the Markov approximation (using both the spectral method and an iterative numerical solution) with those calculated from the full master equation (using both the non-Markovian spectral method and a stochastic simulation).  When $\Delta$ is small there is full agreement between the Markovian and non-Markovian distributions (cf.\ Fig.\ \ref{markov}A, with $\Delta=4.5$); as $\Delta$ grows, the Markovian distributions begin to deviate from the non-Markovian distributions (cf.\ Fig.\ \ref{markov}B, with $\Delta=8.5$).  The deviation is more pronouned at higher $\ell$, i.e.\ for species more downstream in the cascade.  We emphasize, however, that important qualitative features of the distributions, such as modality and locations of the modes, are retained under the approximation.

\subsection{A note on degradation rates}

The ratio $\rho$ of a downstream gene's degradation rate to that of an upstream gene is fixed to 1 in all results in the main text.  Increasing $\rho$ is a computationally straightforward way to obtain, e.g., a bimodal output, since it corresponds to the case in which the downstream gene equilibrates more quickly than the upstream gene, such that the output is simply a weighted sum of distributions peaked at each of the threshold values $q_-$ and $q_+$ (cf.\ Eqn.\ \ref{thresh}).  Specifically, in the $\rho\rightarrow\infty$ limit, for the two-gene cascade $n\xrightarrow{q_n}m$, $p_m = \pi_-e^{-q_-}q_-^m/m! + \pi_+e^{-q_+}q_+^m/m!$, where $\pi_-=\sum_{n\le n_0}p_n$ and $\pi_+=\sum_{n>n_0}p_n$.
However most degradation rates are dominated by cell division, so degradation rate ratios far from than $\sim$1 are unrealistic.  Our results demonstrate that even with all species operating on the same timescale, relatively short cascades with strong enough regulation provide a information-optimal mechanism of converting a unimodal signal to a bimodal signal.

\begin{acknowledgments}
It is a pleasure to acknowledge 
Jake Hofman and
Melanie Lee for their help in this project.
C.W.\ was supported by
NIH 5PN2EY016586-03  and NIH 1U54CA121852-01A1;
C.W.\ and A.M.\ were supported by NSF ECS-0332479;
A.M. was supported by NSF DGE-0742450.
\end{acknowledgments}

\end{document}